\documentclass[nofootinbib,preprint,showpacs,showkeys,amsmath,amssymb]{revtex4}

\usepackage{graphicx} 
\usepackage{dcolumn} 
\usepackage{bm} 
\def\nns{$\sigma_\Delta$\,} 
\def\qqbar{\langle \overline{q}q\rangle} 
\def\lamt{\tilde{\lambda}_\Delta}
\def\lamn{\tilde{\lambda}_N}
\def\pslash{p\!\!\!\slash }

\begin{document}

\title{Meson-baryon sigma terms in QCD Sum Rules}
\author{G. Erkol}
\email{erkol@th.phys.titech.ac.jp}
\author{M. Oka}%
\email{oka@th.phys.titech.ac.jp}
\affiliation{Department of Physics, Tokyo Institute of Technology\\ Meguro, Tokyo 152-8551 Japan}
\begin{abstract}
	We evaluate the pion-nucleon and the pion-Delta sigma terms by employing the method of quantum chromodynamics (QCD) sum rules. The obtained value of the pion-nucleon sigma term is compatible with the larger values already anticipated by the recent calculations. It is also found that the pion-Delta sigma term is as large as the pion-nucleon sigma term.
\end{abstract}

\maketitle
\section{Introduction}
The meson-baryon sigma terms are important for hadron physics as they provide a measure of chiral-symmetry breaking and the scalar quark condensate inside the baryon. In particular, the pion-nucleon sigma term has received much attention and has been extensively analyzed in many problems: in lattice QCD~\cite{Fukugita:1994ba,Dong:1995ec,Gusken:1998wy,Leinweber:2000sa,Procura:2003ig}, Chiral Perturbation Theory ($\chi$PT)~\cite{Bernard:1993nj,Bernard:1996gq,Bernard:1996nu,Borasoy:1995zx,Borasoy:1996bx,Borasoy:1998uu} and various other approaches~\cite{Diakonov:1988mg,Gasser:1990ap,Gasser:1990ce,Jin:1993nn,Hite:2005tg,Lyubovitskij:2000sf,Inoue:2003bk,Schweitzer:2003sb,Schweitzer:2003fg,Chang:2005bq,Flambaum:2005kc}. It is equivalent to the value of the scalar form-factor 
	\begin{equation}\label{sff}
		\hat{m}\langle N(p^\prime)\lvert\overline{u}u+\overline{d}d\rvert N(p) 
		\rangle=\sigma_N(k) 
	\overline{\upsilon}(p^\prime)\upsilon(p),\qquad\hat{m}=\frac{m_u+m_d}{2}, 
	\end{equation}
at zero momentum transfer {\it viz.} $\sigma_{N}=\sigma(0)$, where $\upsilon(p,s)$ is the Dirac spinor for the nucleon, $\hat{m}$ is the average quark mass and $k=(p^\prime-p)^2$. The pion-nucleon sigma term is also defined via the Feynman-Hellmann theorem as 
	\begin{equation}\label{piNsigma} 
		\sigma_N\equiv\sum_{q=u,d}\hat{m}\frac{d\,m_N}{d\,m_q}=\hat{m}\langle 
		N\lvert\overline{u}u+\overline{d}d\rvert N \rangle,
	\end{equation}
where $m_N$ is the nucleon mass. In the chiral limit, the sigma term vanishes; however, the fact that pions and the $u$- and the $d$-quarks are not actually massless implies that $\sigma_N \neq 0$. The pion-nucleon sigma term cannot be measured directly and there is no consensus on its value. Various estimates in the literature range from $18\pm 5$~MeV~\cite{Gusken:1998wy} to $74 \pm 12$~MeV~\cite{Schweitzer:2003fg}.

As for the pion-nucleon sigma term, the pion-Delta sigma term is of recent interest. The Delta resonance, $\Delta(1232)$, plays an important role in pion-nucleon scattering away from threshold and in loop calculations of $\chi$PT~\cite{Jenkins:1991es,Hemmert:1997ye,Bernard:1996gq,Pascalutsa:2002pi,Bernard:2003xf}. It is close in mass to the nucleon and it couples strongly to nucleons, photons and pions. In this respect, the value of the pion-Delta sigma term is crucial in relation to Delta-nucleon mass splitting as the sigma term provides a measure of the shift in the hadron mass away from the chiral limit. It is equivalent to the value of the Delta scalar form-factor at zero momentum transfer {\it viz.} $\sigma_{\Delta}=\sigma_\Delta(0)$, where the latter is defined as
	\begin{equation}\label{Dsff}
		\hat{m}\langle\Delta(p^\prime,s^\prime)\lvert\overline{u}u+\overline{d}d\rvert 
		\Delta(p,s) \rangle=-\overline{\upsilon}_\mu(p^\prime,s^\prime)[g^{\mu\nu} 
		\sigma_\Delta(k) +p^{\prime\nu}p^\mu F_T(k)]\upsilon_\nu(p,s).
	\end{equation}
Here, $\sigma_\Delta(k)$ and $F_T(k)$ are the scalar and tensor form-factors, respectively, and $\upsilon^\mu(p,s)$ is the Rarita-Schwinger spin-vector of the Delta, with the spin projection $s$. The minus sign on the right-hand side (RHS) is conventional like in the case of the free Delta Lagrangian. Analogously to Eq.~\eqref{piNsigma}, the pion-Delta sigma term is defined via the Feynman-Hellmann theorem as 
	\begin{equation}\label{DSdef}
		\sigma_\Delta\equiv\sum_{q=u,d}\hat{m}\frac{d\,m_\Delta}{d\,m_q} 
		=-\hat{m}\langle \Delta(s) \lvert\overline{u}u+\overline{d}d\rvert 
		\Delta(s^\prime) \rangle, 
	\end{equation}
and it corresponds to the change in Delta mass with the quark mass. Recently, some model-dependent approaches have been used to obtain \nns: chiral-quark model produces $\sigma_\Delta=32 \pm 3$~MeV~\cite{Lyubovitskij:2000sf}, which is in agreement with the value from the so-called configuration space quark model, $\sigma_\Delta=30 \pm 2$~MeV~\cite{Cavalcante:2005mb}. A preliminary chiral perturbation theory analysis of the lattice data gives $\sigma_\Delta=20.6$~MeV~\cite{Bernard:2005fy}. A solution of the Faddeev equation produces a relatively larger value as $\sigma_\Delta \simeq 50$~MeV~\cite{Flambaum:2005kc}.

Our aim in this work is to calculate the pion-nucleon and the pion-Delta sigma terms using the external-field QCD sum rules (QCDSR), which are a powerful tool to extract qualitative and quantitative information about hadron properties~\cite{Shifman:1978bx,Shifman:1978by,Reinders:1984sr,Ioffe:1983ju}. In this framework, one starts with a correlation function that is constructed in terms of hadron interpolating fields. On the theoretical side, the correlation function is calculated using the Operator Product Expansion (OPE) in the Euclidian region. This correlation function is matched with an {\em Ansatz} that is introduced in terms of hadronic degrees of freedom on the phenomenological side. The matching provides a determination of hadronic parameters like baryon masses, magnetic moments, coupling constants of hadrons, and so on. To determine the value of the sigma terms, we evaluate the vacuum-to-vacuum transition matrix element of two nucleon and Delta interpolating fields in an external isoscalar-scalar field. Our analysis here follows closely the one in Ref.~\cite{Jin:1993nn}, where $\sigma_N$ was calculated using this method and $\sigma_N=36 \pm 5~\text{MeV}$ was obtained. We depart from this prior analysis by 1) incorporating the dimension-7 operators 2) improving the QCDSR analysis with up-to-date values of the vacuum parameters, in particular the vacuum susceptibilities and 3) treating the uncertainties more systematically by employing a Monte-Carlo analysis. Moreover, we report on the value of the pion-Delta sigma term in QCDSR by following the same approach. 

Our paper is organized as follows: In Section~\ref{secNSR}, we present the
formulation of QCDSR and construct the relevant sum rules. We give the numerical analysis of the sum rules and discuss the results in Section~\ref{secAN}. Finally, we arrive at our
conclusions in Section~\ref{secCONC}.

\section{The derivation of the sum rules}~\label{secNSR}
In the external-field QCDSR method, one starts with the correlation function of the baryon interpolating fields in the presence of an external constant isoscalar-scalar field $S$, defined by 
	\begin{align}\label{cor1a} 
	\begin{split}	
		&i\int d^4 x~ e^{i p\cdot x}\, \left \langle 0\left \lvert{\cal 
		T}[\eta_N(x)\overline{\eta}_N(0)]\right\rvert 0\right\rangle_S 
		=\Pi(p) + S\, \Pi_S (p)+ O(S^2),\\ 
		&i\int d^4 x~ e^{i p\cdot x}\, \left \langle 0\left \lvert{\cal 
		T}[\eta^\mu_\Delta(x)\overline{\eta}^\nu_\Delta(0)]\right\rvert 0\right\rangle_S 
		=\Pi^{\mu\nu}(p) + S\, \Pi^{\mu\nu}_S (p)+ O(S^2),
	\end{split}	
	\end{align}
where $\eta_N$ and $\eta^\mu_\Delta$ are the nucleon and the Delta interpolating fields, respectively:
	\begin{align}\label{intfi} 
	\begin{split}	
		&\eta_N=\epsilon_{abc}\left[u_a^T C\gamma_\mu u_b\right]\gamma_5 \gamma^\mu d_c,\\
		&\eta^\mu_\Delta = \epsilon_{abc}\left[u_a^T C\gamma_\mu u_b\right] u_c .
	\end{split}
		\end{align}
Here $a,b,c$ are the color indices, $T$ denotes transposition and $C=i\gamma^2\gamma^0$. For the interpolating field of nucleon, there are two independent local operators, but the one in Eq.~(\ref{intfi}) is the optimum choice for the lowest-lying positive parity nucleon (see {\em e.g.} Ref~\cite{Jido:1996ia} for a discussion on negative-parity baryons in QCDSR). $\Pi(p)$ and $\Pi^{\mu\nu}(p)$ are the correlation functions when the external field is absent and correspond to the functions that are used to determine the baryon masses. The second terms in Eq.~\eqref{cor1a} represent the linear response of the correlators to a small external scalar-field $S$, which is computed with an additional term to the QCD Lagrangian, which is 
	\begin{equation}\label{addLag} 
		\Delta{\cal L} = -S\,g_q^S\left[\overline{u}(x)\,u(x)\,+ 
		\,\overline{d}(x)\,d(x)\right].
	\end{equation} 
Here, $S$ represents the external scalar-field and $g_q^S$ is associated with the coupling of the external scalar-field to the quark. The external scalar-field contributes to the correlation functions in Eq.~\eqref{cor1a} in two ways: first, it directly couples to the quark field in the baryon currents and second, it modifies the condensates by polarizing the QCD vacuum. In the presence of an external scalar-field there are no correlators that break the Lorentz invariance, like $\langle\overline{q}\sigma_{\mu\nu}q\rangle$ which appears in the case of an external electromagnetic-field $F^{\mu\nu}$. However, the correlators already existing in the vacuum are modified by the external field, {\em viz.}
	\begin{align}\label{vaccon} 
	\begin{split}	
		\qqbar_S {}&\equiv \qqbar - \chi S \qqbar ,\\
		\langle g_c \overline{q} {\bm\sigma}\cdot {\bm G} q\rangle_S {}&\equiv 
		\langle g_c \overline{q} {\bm\sigma}\cdot {\bm G} q\rangle - \chi_G S 
		\langle g_c \overline{q} {\bm\sigma}\cdot {\bm G} q\rangle , 
	\end{split}	
	\end{align}
where $\chi$ and $\chi_G$ are the susceptibilities corresponding to the quark and the quark-gluon mixed condensates, respectively, and the coupling of the external scalar-field to the quark is simply taken as $g_q^S=1$. 

At the quark level, we have 
	\begin{align}\label{cor2}
	\begin{split}	
		\left \langle 0\Big \lvert{\cal T} [\eta_N(x)\overline{\eta}_N(0)] \Big 
		\rvert0\right \rangle_S={}&2 i \epsilon^{abc}\epsilon^{a^\prime b^\prime 
		c^\prime} Tr \{S_u^{b b^\prime}(x) \gamma_\nu C [S_u^{a 
		a^\prime}(x)]^T C \gamma_{\mu}\}\gamma_5\gamma^\mu 
		S_d^{c c^\prime}(x) \gamma^\nu\gamma_5,\\
		\left \langle 0\Big \lvert{\cal T} 
		[\eta_\Delta^\mu(x)\overline{\eta}_\Delta^\nu(0)] \Big 
		\rvert0\right \rangle_S={}&-2 i \epsilon^{abc}\epsilon^{a^\prime b^\prime 
		c^\prime} \Big(Tr \{S_u^{b b^\prime}(x) \gamma^\nu C [S_u^{a 
		a^\prime}(x)]^T C \gamma^{\mu}\}S_u^{c c^\prime}(x)\\
		&\quad +2 S_u^{b b^\prime}(x) \gamma^\nu C [S_u^{a 
		a^\prime}(x)]^T C \gamma^{\mu} S_u^{c c^\prime}(x)\Big). 
	\end{split}
	\end{align}
To calculate the Wilson coefficients, we need the quark propagator $S_q$ in the presence of the external scalar-field, which is given in Ref.~\cite{Jin:1993nn}. Using this quark propagator, one can compute the correlation function $\Pi_S(q)$ and $\Pi^{\mu\nu}_S(q)$, which can be brought into the form
	\begin{align}\label{str} 
	\begin{split}	
		\Pi_S(p)={}&\Pi_N(p^2)\,\pslash+\Pi^\prime_N(p^2),\\
		\Pi^{\mu\nu}_S(p)={}&\Pi_\Delta(p^2)\,g^{\mu\nu}\,\pslash\,+\, 
		\Pi^\prime_\Delta(p^2)\,g^{\mu\nu} +\dotsb,
	\end{split}	
	\end{align}
where the ellipsis represents the Lorentz-Dirac structures other than $g_{\mu\nu}$ and $g_{\mu\nu}\pslash$.

Note that one can obtain the sum rules at different Lorentz structures. Here, we choose to work with the sum rules at the structures $\pslash$ and $g^{\mu\nu}\,\pslash$ for the nucleon and the Delta, respectively, where the latter is completely contributed by the Delta baryons with $J=\frac{3}{2}$ (see {\it e.g.} Ref.~\cite{Ioffe:1983ju} for details). The OPE sides of the sum rules linear in $S$ are then obtained as 
	\begin{align}
	\begin{split}	
		\label{qmomN} \Pi_N(p^2)=&{}\frac{S}{(2\pi)^4}\left [\,\,a\, 
		\ln(-p^2)+\frac{4}{3p^2}\kappa a^2\,\chi + \frac{m_0^2}{2p^2}a 
		+\frac{m_0^2}{6 p^4}a^2(\chi+\chi_G)\right],\\
		\Pi_\Delta(p^2)=&{}\frac{S}{(2\pi)^4}\left [\,3\,a\, 
		\ln(-p^2)-\frac{8}{3p^2}\kappa a^2\,\chi -\frac{3 m_0^2}{2p^2}a 
		-\frac{7m_0^2}{9p^4}a^2(\chi+\chi_G)\right].
	\end{split}	
	\end{align}
In the above equations, we have defined the quark condensate $a=-(2\pi)^2\langle\overline{q}q\rangle$, and the quark-gluon--mixed condensate $\langle\overline{q}g_c\sigma \cdot G q\rangle=m_0^2 \langle\overline{q}q\rangle$ with the QCD coupling-constant squared $g_c^2=4\pi\alpha_s$. The four-quark condensate is parameterized as $\langle(\overline{q}q)^2 \rangle\equiv\kappa \langle\overline{q}q\rangle^2$. We note that the last term of $\Pi_N(p^2)$, which is associated with the dimension-7 operators has been neglected in Ref.~\cite{Jin:1993nn}.

The analyticity of the correlation function allows us to write the phenomenological side of the sum rules in terms of a double-dispersion relation of the form
	\begin{equation}\label{phenside}
		\text{Re}\,\Pi_B(p^2)=\frac{1}{\pi^2}\int^\infty_0 \int^\infty_0 
		\frac{\text{Im}\,\Pi_B(p)}{(s_1-p^2)(s_2-p^2)}\,ds_1\,ds_2.
	\end{equation}
The ground-state hadron contribution is singled out by utilizing the zero-width approximation, where the hadronic contributions from the Breit-Wigner form to the imaginary part of the correlation function is proportional to the $\delta$-function:
	\begin{equation}~\label{satN}
	\begin{split}	
		\!\!\!&\text{Im}\,\Pi_N(p)={}\pi^2\delta(s_1-m_N^2) \delta(s_2-m_N^2) \langle 0\lvert\eta_N \rvert
		N(p) \rangle \langle 	
		N(p)\lvert S(\overline{u}u+\overline{d}d) \rvert N(p)\rangle 
		\langle N(p) \lvert \overline{\eta}_N \rvert 
		0\rangle\\
		\!\!\!&\quad+\pi^2\delta(s_1-m_N^2) \delta(s_2-m_{N^\ast}^2)\langle 0\lvert\eta_{N} \rvert
		N(p) \rangle \langle 
		N(p)\lvert S(\overline{u}u+\overline{d}d) \rvert N^\ast(p)\rangle 
		\langle N^\ast(p) \lvert \overline{\eta}_N \rvert 
		0\rangle.
	\end{split}	
	\end{equation}
Here, the second term is associated with the transitions to higher nucleon states. We then expresses the correlation function for the nucleon as a sharp resonance plus a continuum after Borel transformation:
	\begin{equation}\label{phensideN}
		\Pi_N(M^2)=\left(\lambda_N^2 m_N 
		\frac{\sigma_N}{\hat{m}}+\,C_N\,M^2\right)\, 
		\,\frac{e^{-m_N^2/M^2}}{M^4}+\frac{1}{\pi} 
		\int_{w^2}^\infty\,\frac{\text{Im} 
		\,\Pi_N}{M^4}\, e^{-t/M^2}\,dt,
	\end{equation}
by using the definition in Eq.\eqref{piNsigma}. The correlation function for the Delta is similarly expressed as
	\begin{equation}\label{phensideD}
		\Pi_\Delta(M^2)=-\left(\lambda_\Delta^2 m_\Delta 
		\frac{\sigma_\Delta}{\hat{m}}+\,C_\Delta\,M^2\right)\,
		\,\frac{e^{-m_\Delta^2/M^2}}{M^4}+\frac{1}{\pi} 
		\int_{w^2}^\infty\,\frac{\text{Im} 
		\,\Pi_\Delta}{M^4}\, e^{-t/M^2}\,dt,
	\end{equation}
In Eqs.~\eqref{phensideN} and \eqref{phensideD}, the matrix elements of the currents $\eta_N$ and $\eta^\mu_\Delta$ between the vacuum and the hadron states are defined as 
	\begin{align}~\label{overlapN}
	\begin{split}	 
		\langle 0 \lvert \eta_N \rvert N(p,s) 
		\rangle={}& \lambda_N \upsilon(p,s),\\
		\langle 0 \lvert \eta^\mu_\Delta \rvert \Delta(p,s) 
		\rangle={}& \lambda_\Delta \upsilon^\mu(p,s), 
	\end{split}	
	\end{align}
respectively, for the nucleon and the Delta, where $\lambda_N$ and $\lambda_\Delta$ are the residues. We also make use of the Rarita-Schwinger spin-sum, which is 
	\begin{equation}
		~\label{RSss} \sum_s \upsilon^\mu(p,s) \overline{\upsilon}^\nu(p,s)=-\left( 
		g^{\mu\nu}-\frac{1}{3}\gamma^\mu \gamma^\nu-\frac{p^\mu 
		\gamma^\nu-p^\nu\gamma^\mu}{3\,m_\Delta}-\frac{2\,p^\mu 
		p^\nu}{3\,m_\Delta^2}\right)(\pslash+m_\Delta). 
	\end{equation}
We have included the single-pole contributions with the factors $C_N$ and $C_\Delta$, which correspond to the transitions to higher baryon states. These transition terms are not properly suppressed after the Borel transformation and should be included on the phenomenological side.

The QCD sum rules are obtained by matching the OPE sides with the hadronic sides and applying the Borel transformation. The resulting sum rules are 
	\begin{align}\label{sumq}
	\begin{split}	
		&-a\,M^4\,E_0\,-\,\frac{4 M^2}{3} \, 
		\chi\,\kappa\, a^2\, L^{4/9}\,- 
		\,\frac{m_0^2}{2}\,M^2\,a\,L^{-14/27}\,+\frac{m_0^2}{6}\, 
		a^2\,(\chi+\chi_G)\,L^{-2/27}\,\\
		&\quad =\left(\lamn^2 m_N
		\frac{\sigma_N}{\hat{m}}+\,C_N\,M^2+\, 
		\frac{M^2}{2} w^4 \,\delta w^2\,L^{-4/9} 
		e^{(m_N^2-w^2)/M^2}\right) \,e^{-m_N^2/M^2},
	\end{split}\\
	\begin{split}\label{sumqD}
		&3 a\,M^4\,E_0\,L^{16/27}\,-\,\frac{8M^2}{3} \, 
		\chi\,\kappa\, a^2\, L^{28/27}\,- 
		\,\frac{m_0^2}{2}\,M^2\,a\,L^{2/27}+\,\frac{7m_0^2}{9}\, 
		a^2\,(\chi+\chi_G)\,L^{14/27}\,\\
		&\quad =\left(\lamt^2 m_\Delta 
		\frac{\sigma_\Delta}{\hat{m}}+\,C_\Delta\,M^2 
		+\, \frac{M^2}{5} w^4\,\delta w^2\,L^{4/27} 
		e^{(m_\Delta^2-w^2)/M^2}\right)\,e^{-m_\Delta^2/M^2},
	\end{split}
	\end{align}
where $M$ is the Borel mass and we have defined $\tilde{\lambda}_B^2=32 \pi^4 \lambda_B^2$ with $B=N,~\Delta$. The continuum contributions are included via the factor 
	\begin{equation}
		E_n\equiv 1- (1+x+...+\frac{x^n}{n!})e^{-x}\, , 
	\end{equation}
with $x=w^2/M^2$, where $w$ is the continuum threshold. In the sum rules above,  the third terms on the RHS of the sum rules denote the contributions that come from the response of the continuum threshold to the external field. Here, $\delta w^2$ represents the variation of the continuum threshold and the coefficient is calculated by differentiating the continuum parts of the chiral-even nucleon and Delta mass sum rules~\cite{Ioffe:1981kw} with respect to the quark mass. These terms are suppressed by the factor $e^{-(w^2-m_B^2)/M^2}$ as compared to the single-pole terms, however, should be included on the phenomenological side if $\delta w^2$ is large (see Ref.~\cite{Ioffe:1995jt} for a detailed explanation of this term). The corrections that come from the anomalous dimensions of various operators are included with the factors $L=\log(M^2/\Lambda_{QCD}^2)/\log(\mu^2/\Lambda_{QCD}^2)$, where $\mu=500$~MeV is the renormalization scale and $\Lambda_{QCD}$ is the QCD scale parameter.

\section{Analysis of the sum rules}~\label{secAN}
We determine the uncertainties in the extracted parameters via the Monte-Carlo based analysis introduced in Ref.~\cite{Leinweber:1995fn}. In this analysis, randomly selected, Gaussianly distributed sets are generated from the uncertainties in the QCD input parameters. Here we use $a=0.52 \pm 0.05$~GeV$^3$, $b\equiv\left\langle g_c^2 G^2\right\rangle=1.2 \pm 0.6$~GeV$^4$, $m_0^2=0.72 \pm 0.08$~GeV$^2$, and $\Lambda_{QCD}=0.15 \pm 0.04$~GeV. The factorization violation in the four-quark operator is searched via the parameter $\kappa$, where we take $\kappa=2 \pm 1$ and $1 \leq \kappa \leq 4$; here $\langle (\overline{q}q)^2\rangle \geq \langle \overline{q}q\rangle^2$ is assumed via the cut-off at 1 (for a discussion on QCD parameters see {\it e.g.} Ref.~\cite{Leinweber:1995fn}). The value of the susceptibility $\chi$ has been calculated in Ref.~\cite{Erkol:2005jz} as $\chi= -10 \pm 1 ~\text{GeV}^{-1}$ and has been used to calculate the scalar-meson--baryon coupling constants~\cite{Erkol:2006eq}. The susceptibility $\chi_G$ is less certain, therefore we take its value equal to the one of $\chi$, however with a larger uncertainty $\chi_G=-10 \pm 3 ~\text{GeV}^{-1}$. We use $10^4$ such configurations from which the uncertainty estimates in the extracted parameters are obtained using a fit of the LHS of the sum rules to the RHS. For $\hat{m}$, we make use of the Gell-Mann--Oakes--Renner relation which is 
	\begin{equation}
		\label{GOR} 2\hat{m}\langle\overline{q}q\rangle = -m_\pi^2 f_\pi^2,
	\end{equation}
where $m_\pi=138~\text{MeV}$ is the pion mass and $f_\pi=93~\text{MeV}$ is the pion decay constant. We use the following chiral-odd nucleon and Delta mass sum rules for normalization of the sigma-term sum rules~\cite{Belyaev:1982sa,Hwang:1994vp}: 
		\begin{gather}~\label{nmass1}
			a\,M^4-\frac{5}{72}a\,b=\, 
			\frac{\lamn^2}{2}m_N\,e^{-m_N^2/M^2},\\
			~\label{deltamass1} 
			\frac{4}{3}a\,E_1\,L^{16/27}M^4-\frac{2}{3}E_0\,m_0^2\, a\, 
			L^{2/27}M^2 -\frac{1}{18} a\, b\, L^{16/27}= 
			\,\frac{\lamt^2}{2}m_\Delta\,e^{-m_\Delta^2/M^2} ,
		\end{gather}
which have been found to be more reliable than the chiral-even sum rules~\cite{Leinweber:1995fn,Lee:1997ix}. The Monte-Carlo analysis of the sum rules \eqref{sumq} and \eqref{sumqD} is performed by first fitting the mass sum rules to obtain the pole residues $\tilde{\lambda}_B$ and these residue values are used in the sigma-term sum rules for each corresponding parameter set. We find that the nucleon mass sum rule in Eq.~\eqref{nmass1} fails to resolve the pole from the continuum. In order to proceed and reduce the uncertainties in the final results as much as possible, we fix the nucleon mass at its experimental value as $m_N=0.94$~GeV and the continuum threshold at $w=1.5$~GeV. This produces $\tilde{\lambda}_N^2=1.60 \pm 0.18$~GeV$^6$. It is also found that the mass sum rule \eqref{deltamass1} somewhat overestimates the value of the Delta mass~\cite{Lee:1997ix}. Therefore, instead of fixing the mass at its experimental value, we make a two parameter fit including $m_\Delta$ and $\tilde{\lambda}_\Delta$ by fixing the continuum threshold at $w=1.7$~GeV, which gives $m_\Delta=1.45 \pm 0.05$~GeV and $\tilde{\lambda}_\Delta^2=5.20 \pm 0.66$~GeV$^6$.

	\begin{figure}
		[th] 
		\includegraphics[scale=0.40]{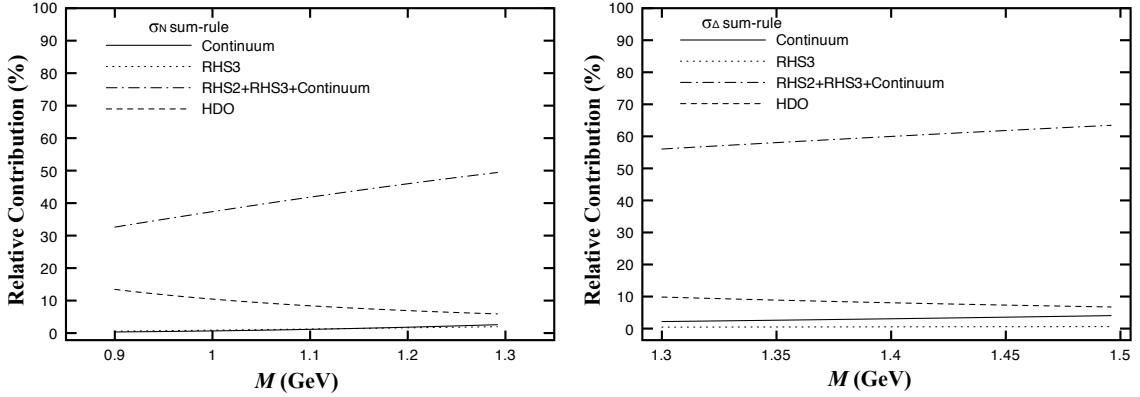} \caption{The continuum-plus-transition contributions as well as those of continuum change for the sum rules~\eqref{sumq} and~\eqref{sumqD}.} \label{contrN}
	\end{figure}

The valid Borel regions are determined so that the highest-dimensional operator (HDO) contributes no more than about $10\%$ to the OPE side which gives the lower limit on the valid Borel region. The upper limit is determined using a criterion such that the continuum contributions are less than $50\%$ of the phenomenological side. It is relevant to point out that the dominant contributions to the OPE sides of the sum rules in \eqref{sumq} and \eqref{sumqD} come from the terms that involve the susceptibilities \emph{i.e.} the second and the fourth terms on the LHS. This leads to a suppression of the continuum contributions, which are involved by the first terms on the LHS only. In Figure~\ref{contrN}, we plot the continuum-plus-transition contributions as well as those of continuum change for the sum rule~\eqref{sumq}. In the Borel window $0.9$~GeV$\leq M\leq 1.3$~GeV, which we determine as the fiducial region, the relative double-pole-contribution is higher than $50\%$, while the continuum contributes less than about $10\%$. We also observe that the effects of the continuum variation with the external field are minor. For the sum rule \eqref{sumqD}, we are not able to find a Borel window according to the above criterion: the double-pole contribution is less than $50\%$ in the region where the HDO starts to contribute less than $10\%$. As can be observed in Figure~\ref{contrN}, in the Borel window $1.3$~GeV$\leq M\leq 1.5$~GeV which we take as the fiducial region for the sum rule analysis, contribution of the continuum-plus-transition to the RHS of the sum rule is about $60\%$, while the HDO contributes less than $10\%$ to the OPE side.

\begin{figure}
		[th] 
		\includegraphics[scale=0.40]{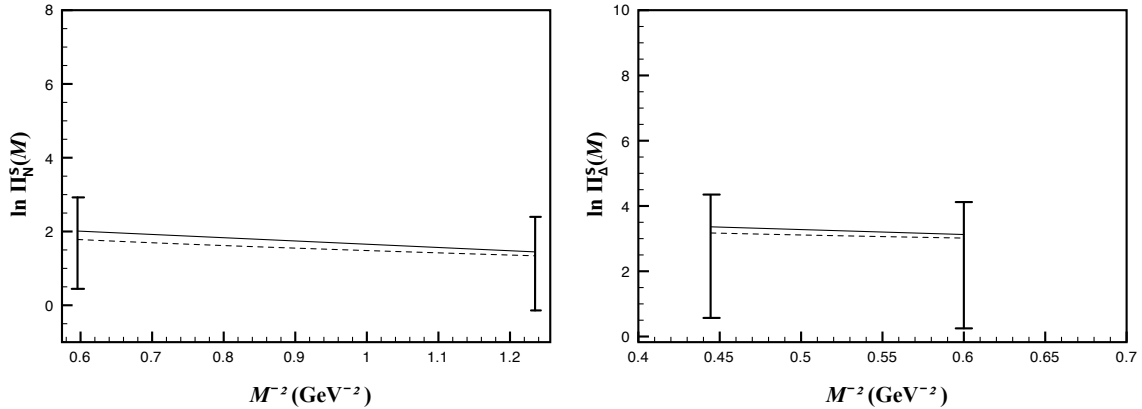} \caption{The subtracted form of the sum rules~\eqref{sumq} and~\eqref{sumqD}. The solid line is the double-pole contribution and the dashed-line is the OPE-minus-excited states and minus-continuum-change contributions, where we use the average values of the QCD and the obtained fit parameters. The error bars at the two ends represent the uncertainties in the QCD parameters.} \label{logN}
	\end{figure}

To demonstrate how well the sum rules and the fitting work, we first arrange the sum rules in the subtracted form 
	\begin{align}
		\begin{split} 
			\Pi_N^s &\equiv \tilde{\lambda}_N^2 m_N\frac{\sigma_N}{\hat{m}} e^{-m_N^2/M^2},\\
			\Pi_\Delta^s &\equiv \tilde{\lambda}_\Delta^2 m_\Delta\frac{\sigma_\Delta}{\hat{m}} e^{-m_\Delta^2/M^2},
		\end{split}
	\end{align} 
where $\Pi_B^s$ represents the OPE-minus-excited and minus-continuum-change contributions. As the RHS appears as a straight line with this form, the linearity of the LHS gives an indication of the OPE convergence and the quality of the continuum model. In Figure~\ref{logN}, we present the fit of the sum rules \eqref{sumq} and \eqref{sumqD}, respectively, using the average values of the QCD and the obtained fit parameters. The error bars at the two ends correspond to the uncertainties in the QCD parameters.

Since the continuum contributions in the sum rules \eqref{sumq} and \eqref{sumqD} are suppressed as compared to the total phenomenological side, it becomes difficult to extract information about the continuum threshold from the fit. Therefore, we have assumed that the continuum thresholds are equivalent to those for the mass sum rules. The variation of the continuum threshold $\delta w^2$ can also be obtained from the fit. However, instead of taking this as a free parameter, we proceed with a generous assumption that the continuum threshold changes by $25\%$ with the external field \emph{viz.} $\delta w^2=w/4$. Then, a two parameter fit of the sum rule \eqref{sumq} including the $\sigma_N$ and $C_N$ from a consideration of $10^4$ parameter sets produces
	\begin{eqnarray}\label{sigmaN}
		\sigma_N=53\pm 24~\text{MeV},
	\end{eqnarray}
and a fit of the the sum rule \eqref{sumqD} including the $\sigma_\Delta$ and $C_\Delta$ gives
	\begin{eqnarray}\label{sigmaD}
		\sigma_\Delta=54\pm 25~\text{MeV}.
	\end{eqnarray}
Eq.~\eqref{sigmaN} suggests a $60\%$ enhancement of the pion-nucleon sigma term as compared to the value obtained in Ref.~\cite{Jin:1993nn}, which is $36 \pm 5~\text{MeV}$. The large errors of about $50\%$ in our results (as compared to typical $30\%$ in QCDSR) mainly stem from the uncertainties in the residue values and the vacuum susceptibilities. We also find that the dimension-7 terms contribute by $20\%$ to the OPE side. As emphasized above, the continuum effects are suppressed, therefore a change in the value of the continuum threshold \emph{e.g.} by $10\%$, leads to a negligible change in the final values. One can also obtain a ratio of the pion-nucleon sigma term to the pion-Delta sigma term by dividing the corresponding values of the two sigma terms at each QCD parameter set. The analysis of the final distribution obtained from this method produces
	\begin{equation}\label{rat}
			\sigma_N/\sigma_\Delta=1.06 \pm 0.16,
	\end{equation}
which is in accordance with the observation that the pion-Delta sigma term is as large as the pion-nucleon sigma term. 

\section{Conclusions and discussion}~\label{secCONC}
We have calculated the pion-nucleon and the pion-Delta sigma terms using the external-field QCDSR method. Our analysis includes several improvements over the prior work in Ref.~\cite{Jin:1993nn}, which enhance the central value of the pion-nucleon sigma term by $60\%$. We observe that inclusion of the dimension-7 operators in the OPE analysis brings a contribution of $20\%$ to the final results. A Monte-Carlo analysis of the QCD input parameters provides a more systematic treatment of the errors. Although the errors in the final results are large (which could be best improved by a more precise determination of the residues), a number of qualitative features are evident. The value we have obtained for the pion-nucleon sigma term is consistent with the larger values already anticipated by the recent calculations. Moreover, our analysis favors a pion-Delta sigma term as large as the pion-nucleon sigma term. 

In Ref.~\cite{Griegel:1994xb}, inconsistencies were pointed out in the leading nonanalytic behavior in $m_q$ between the treatment of the nucleon mass with the QCDSR and the $\chi$PT description. The inconsistency arises because the quark condensate has a chiral behavior of the form
	 \begin{equation}\label{chlog}
	 	\qqbar= \mathring{\langle\overline{q}q\rangle} \Big(1-\frac{3}{32\pi^2 \mathring{f}_\pi^2}\,\mathring{m}_\pi^2 \ln \frac{\mathring{m}_\pi^2}{M_0^2}+\ldots\Big),
	 \end{equation}
whereas the nucleon mass is known to have a leading nonanalytic term proportional to $m_q^{3/2}$ in the form
	\begin{equation}\label{msterm}
	M_N=\mathring{M}_N + A\, \mathring{m}_\pi^2-\frac{3 \mathring{g}_A^2}{32\pi \mathring{f}_\pi^2}\, \mathring{m}_\pi^3+\ldots,
	\end{equation}
but no $m_q\ln m_q$ term. (Here $g_A$ is the nucleon axial-vector coupling constant, $M_0$ is a mass parameter of the order of $\sim 1$~GeV, $A$ is an unknown constant~\cite{Gasser:1987rb} and all quantities with a circle over denote the first term in the chiral expansion of that quantity.) These spurious and missing nonanalytic terms lead to an intrinsic uncertainty of the order of $\sim 100$~MeV in the QCDSR estimates of $M_N$ and $\sigma_N$. It was shown in Ref.~\cite{Lee:1994hs} with a more general argument that the spurious chiral-log contributions to $M_N$ originate from virtual pions, which cancel if the continuum contribution to the correlator is treated carefully by including the $\pi$-N continuum via soft-pion theorem. Unlike the case of $M_N$, it is not straightforward to prove the same for the current sum rule for $\sigma_N$, by explicitly including the $\pi$-N continuum. The sum rule in Eq.~(\ref{sumq}) takes account of the change of $w$ with $m_q$ in a complicated way. As stated in Ref.~\cite{Griegel:1994xb}, this $m_q$ dependence of $w$ may eliminate the discrepancy between the QCDSR and the $\chi$PT descriptions. One should, however, bear in mind that the usual continuum model is inconsistent with the treatment of the continuum by including the virtual pions~\cite{Lee:1994hs}, and difficulties arise with such an arbitrary solution. On the other hand, our approach is equivalent to evaluating the derivative in the Feynman-Hellmann theorem using a QCDSR estimate of $M_N$, which yields a QCDSR estimate of $\sigma_N$. Such a relation between the sum rule of $M_N$ and that of $\sigma_N$ implies that spurious chiral-log contributions to $\sigma_N$ should also cancel with the inclusion of the $\pi$-N continuum. However, we note that, although the spurious terms and the resulting uncertainties are removed by a proper treatment of the continuum, the problem of the missing $m_q^{3/2}$ term, which originates from the chiral expansion of $M_N$ in Eq.~(\ref{msterm}), persists. We know of no satisfactory method for recovering this missing term, which is left as an open problem in QCDSR. The lack of this term leads to uncertainties of $\sim 15$~MeV and $\sim 20$~MeV for $M_N$ and $\sigma_N$ in QCDSR~\cite{Griegel:1994xb}. 

The sigma term is the measure of the contribution of explicit chiral-symmetry breaking in the baryon masses. The QCD Hamiltonian consists of the chiral-invariant terms containing the gauge couplings of gluons and the chiral--non-invariant quark-mass term. Suppose that the chiral
non-invariant term is weak and therefore treated perturbatively. Then the sigma term is nothing but the contribution of the quark-mass term to the baryon mass. As we find that the pion-Delta sigma term is of similar size with the pion-nucleon sigma term, the quark-mass term is concluded to give little contribution to the Delta-nucleon mass difference. This in turn means that their mass difference is entirely due to the gluon-quark gauge-coupling term. Its main role is to induce the spontaneous chiral-symmetry breaking, giving the quark constituent mass, but the perturbative part is known to yield spin-spin (color-magnetic) interaction between the quarks. The latter is well-known to cause the Delta-nucleon mass difference in the quark model.

Finally, we note that the central values of the sigma terms obtained in Eqs.~\eqref{sigmaN} and~\eqref{sigmaD} correspond to the $6\%$ and $4\%$ of the physical nucleon- and the Delta-masses, respectively, which would indicate the change in these baryon masses if the chiral symmetry is restored. We plan to extend the QCDSR analysis to hyperons and calculate the hyperon sigma terms, as well, to address the explicit chiral-symmetry breaking in the complete baryon octet~\cite{erkol_turan_oka}.
\acknowledgments
Discussions with G\"{u}rsevil Turan are gratefully acknowledged. This work has been supported by the Japan Society for the Promotion of Science under contract number P06327.
\bibliography{st}

\end{document}